\documentstyle[12pt,titlepage]{article}
\newlength{\extraspace}
\newlength{\extraspaces}
\newif\ifnoepsf 

\ifnoepsf\else\input{epsf.sty}\fi

\newcommand{\figinclude}[2]%
{\ifnoepsf\fbox{figure {\tt #2} goes here}%
\else{\epsfxsize=#1 pt \epsfbox{#2}}\fi}
\newcommand{\be}{\begin{equation}
\addtolength{\abovedisplayskip}{\extraspaces}
\addtolength{\belowdisplayskip}{\extraspaces}
\addtolength{\abovedisplayshortskip}{\extraspace}
\addtolength{\belowdisplayshortskip}{\extraspace}}
\newcommand{\ee}{\end{equation}}
\newcommand{\ba}{\begin{eqnarray}
\addtolength{\abovedisplayskip}{\extraspaces}
\addtolength{\belowdisplayskip}{\extraspaces}
\addtolength{\abovedisplayshortskip}{\extraspace}
\addtolength{\belowdisplayshortskip}{\extraspace}}
\newcommand{\ea}{\end{eqnarray}}

\newcommand{\tr}{\, {\rm tr} \,}

\newcommand{\N}{\bf N}

\newcommand{\OO}{O}

\newcommand{\unity}{{\bf 1}}

\newcommand{\as}{\alpha_{\rm s}}

\newcommand{\phip}{\phi_+}
\newcommand{\phim}{\phi_-}

\newcommand{\la}{\left\langle}
\newcommand{\ra}{\right\rangle}

\newcommand{\beq}{\begin{equation}}
\newcommand{\eeq}{\end{equation}}
\newcommand{\bea}{\begin{eqnarray}}
\newcommand{\eea}{\end{eqnarray}}
\newcommand{\E}{{\rm e}}

\newcommand{\lt}{\left(}
\newcommand{\rt}{\right)}

\newcommand{\oon}{\frac{1}{N}}

\newcommand{\pa}{\partial}

\begin{document}
\thispagestyle{empty}
\begin{flushright}
SPhT94/110\\
NUP-A-94-17 \\
TIT/HEP--270\\
hep-th/9409157 \\
September, 1994
\end{flushright}
\vspace{13mm}
\begin{center}
{\large{\bf{ 
LARGE N RENORMALIZATION GROUP APPROACH TO MATRIX MODELS 
}}}\footnote{A talk given at the XXth International Colloquium on 
Group Theoretical Methods in Physics,\\
\indent \ \ Toyonaka, July 4-9, 1994} \\[18mm]
{\sc Saburo Higuchi}
\footnote{e-mail: {\tt higuchi@amoco.saclay.cea.fr}}   
\\
{\it Service de Physique Th\'eorique, \\
 CEA Saclay, 91191 \mbox{Gif-sur-Yvette} CEDEX, France}, \\[2mm]   
{\sc Chigak Itoi}
\footnote{e-mail: {\tt itoi@phys.cst.nihon-u.ac.jp}}
\\
{\it Department of Physics and Atomic Energy Research Institute, \\
College of Science and Technology, Nihon University, \\
Kanda Surugadai, Chiyoda, Tokyo 101, Japan}, \\[2mm]
{\sc Shinsuke Nishigaki}
\footnote{e-mail: {\tt nsgk@th.phys.titech.ac.jp} \\
\indent \ \ address after 1 October 1994:  Department of Physics,
Technion, Haifa 32000, Israel} 
 and  
{\sc Norisuke Sakai}
\footnote{e-mail: {\tt nsakai@th.phys.titech.ac.jp},\ \ Speaker}
 \\
{\it Department of Physics, Tokyo Institute of Technology, \\
Oh-okayama, Meguro, Tokyo 152, Japan} \\[5mm]

\ \\

\ \\

{\bf Abstract}\\[7mm]
{\parbox{13cm}{\hspace{5mm}
We summarize our 
recent results 
on the large $N$ renormalization group (RG) approach to 
matrix models for discretized two-dimensional quantum gravity. 
We derive exact RG equations by solving the 
reparametrization identities, which reduce 
infinitely many induced interactions to a finite number of them. 
We find a nonlinear RG equation and an algorithm to 
obtain the fixed points and the scaling exponents. 
They reproduce  the spectrum 
of relevant operators in
the exact solution. 
The RG flow is visualized by the linear approximation.
}}
\end{center}
\setcounter{footnote}{0}
\newpage
\setcounter{page}{1}
\setlength{\baselineskip}{14pt}
\vspace{6mm}
\begin{center}
{\large{\bf{ 
LARGE N RENORMALIZATION GROUP APPROACH TO MATRIX MODELS 
}}}
 \\[4mm]
\begin{small}
{\sc  
 Saburo Higuchi} 
\\
{\it Service de Physique Th\'eorique, \\
 CEA Saclay, 91191 \mbox{Gif-sur-Yvette} CEDEX, France}, \\[1mm]   
{\sc  
 Chigak Itoi} 
\\
{\it Department of Physics and Atomic Energy Research Institute, \\
College of Science and Technology, Nihon University, \\
Kanda Surugadai, Chiyoda, Tokyo 101, Japan}, \\[1mm]
{\sc  
 Shinsuke Nishigaki} 
\footnote{
Address after 1 October 1994:  Department of Physics,
Technion,
Haifa 32000, 
Israel},
and 
{\sc  
Norisuke Sakai} 
\footnote{Speaker}
 \\
{\it Department of Physics, Tokyo Institute of Technology, \\
Oh-okayama, Meguro, Tokyo 152, Japan} \\[2mm]
{\bf Abstract}\\[1mm]
{\parbox{13cm}{
We summarize our 
recent results 
on the large $N$ renormalization group (RG) approach to 
matrix models for discretized two-dimensional quantum gravity. 
We derive exact RG equations by solving the 
reparametrization identities, which reduce 
infinitely many induced interactions to a finite number of them. 
We find a nonlinear RG equation and an algorithm to 
obtain the fixed points and the scaling exponents. 
They reproduce  the spectrum 
of relevant operators in
the exact solution. 
The RG flow is visualized by the linear approximation.
}}
\end{small}
\end{center}

\setcounter{equation}{0}

\noindent
{\bf 1. Introduction}\\
Matrix models have been a powerful tool to study 
two-dimensional quantum gravity (2D QG)
via simplicial decomposition of the spacetime${}^1$. 
We are particularly interested in 
the 2D QG
both as a string theory and a toy model for the QG in higher 
dimensions. 
Exact solutions of the matrix model 
have been successfully obtained for 2D QG coupled to 
minimal conformal matter with central charge $c \le 1$,
but unsuccessful for $c > 1$ cases which correspond to 
realistic string theories. 
We need to obtain approximation schemes which 
provide us with correct results for the 
exactly solved cases and 
enable us to calculate critical coupling constants and 
critical exponents for unsolved matrix models, 
especially for $c>1$.

A large $N$ 
renormalization group${}^2$ (RG)
has been proposed 
by Br\'{e}zin and Zinn-Justin 
as such an approximation method${}^3$, 
identifying the size of the matrix with
the inverse lattice constant of the simplices,
and was discussed by several groups${}^4$. 
We have succeeded to derive explicitly exact RG 
equations for $O(N)$-vector models${}^5$
and one-${}^6$ and two-matrix models${}^7$.
We found that it is crucial to take account of the reparametrization 
identities in order to obtain a meaningful RG 
equation. 
The RG equation for the 
matrix model turned out to be 
nonlinear in contrast to the linear RG 
equation for the vector model. 
Moreover, we found that the 
global picture of the RG flow can be drawn practically 
by a linear approximation to the nonlinear 
RG equation${}^7$. \\

\noindent
{\bf {2. Nonlinear RG equation for the one-matrix model}}\\
The partition function $Z_N(g_j)$ of a one-matrix model with a
general potential $V(\phi) = \sum_{k\geq 1} \frac{g_k}{k} \phi^k$ 
is defined by an integral over an
$N\times N$ hermitian matrix $\phi$,
$ Z_N(g_j) =   \int d \phi \ \E^{-N  \tr V(\phi)}$.
Starting from 
an $(N+1) \times (N+1)$ hermitian matrix variable $\Phi$,  
we decompose it 
into an $N \times N$ hermitian matrix $\phi$,
a complex $N$-vector $v$
and a real scalar $\alpha$.
Taking cubic interaction 
$V(\phi)={1 \over 2}\phi^2+{g \over 3}\phi^3$, 
 we integrate over $v$ exactly to obtain
\begin{equation}
Z_{N+1}(g) 
= 
\left(\frac{\pi}{N+1}\right)^N
\int d\phi \ 
  {\rm e}^{ -(N+1) \tr V(\phi)} 
\cdot \int d\alpha  \ 
 {\rm e}^{ -(N+1) V(\alpha) - \tr \log (\unity + g (\phi +
  \alpha 
)) } .
\end{equation}
We can evaluate the $\alpha$-integral by the saddle point method,
systematically in  
$1/N$-expansion.
If we explicitly retain only the 
leading part of the $1/N$-expansion, 
we obtain a difference equation relating $Z_{N+1}$ and $Z_N$. 
We shall denote 
the average 
$
Z_N^{-1} \int d\phi (\cdots) \E^{-N \tr V(\phi)} 
$
by $\langle\cdots\rangle$. 
Factorization holds 
in the large-$N$ limit
$  \langle {\cal O}\ {\cal O}' \rangle = 
  \langle {\cal O} \rangle \langle {\cal O}' \rangle + \OO(N^{-2})
$
for a multi-point function of $U(N)$-invariants  
${\cal O},\ {\cal O}'$.
By using this property we obtain for the free energy
$F(N,g)=$\\
$-N^{-2}\log (Z_N(g)/Z_N(0))$ 
\begin{equation}
\left[ N\frac{\partial}{\partial N}+2 \right]  F(N,g) 
=
- \frac 12 + \la \frac 1N  \tr V(\phi) \ra + 
V( \langle \as \rangle) +
\la \frac 1N \tr \log \left( \unity + g(\phi + \la \as \ra 
) 
\right) \ra
  + O \lt \oon \rt.
\label{2}
\end{equation}
Here $\alpha_{\rm s}=\alpha_{\rm s}(\phi)$ 
is determined 
as a $U(N)$-invariant function by the saddle point equation
\beq
V'(\alpha_{\rm s})+\oon \tr \frac{1}{(1/g+\alpha_{\rm s})+\phi}=0.
\eeq

The right hand side of eq.~(\ref{2}) consists of products of
$\langle \oon \tr \phi^j \rangle =j \pa F/\pa g_j$, $j=1,2,\ldots$.
The crucial observation we make is that there is an ambiguity to define 
a flow in the coupling constant space, 
since we can make an arbitrary reparametrization 
of matrix variable $\phi$ in the partition function. 
This will enable us to express 
the $\partial F/\partial g_j$-dependence $(j\neq 3)$
in terms of $\partial F/\partial g_3$, and 
to reduce all the induced interactions to those in 
the original potential. 
The resulting identities are a part of the discrete 
Schwinger-Dyson equation for the system. 

More concretely, 
if we perform reparametrizations regular at the origin,
$  \phi' = \phi + \epsilon \phi^{n+1} \ (n \geq -1) $
for the partition function, 
we obtain identities
\beq
  \sum_{j=0}^{n} 
  \left\langle   \frac{1}{N} \tr \phi^j
  \frac{1}{N} \tr \phi^{n-j} \right\rangle 
  =
  \left\langle 
  \frac{1}{N} \tr\left( \phi^{n+1} V'(\phi) \right)
\right\rangle \ \ (n \geq -1).
\label{eqn:sd-eq}
\eeq
These identities imply relationship among $\pa F/\pa g_j$'s
in the large $N$ limit,
\beq
2n\frac{\pa F}{\pa g_n}+
\sum_{j=1}^{n-1} j(n-j)\frac{\pa F}{\pa g_j} 
\frac{\pa F}{\pa g_{n-j}}=
\sum_{k \geq 1} (n+k) g_k \frac{\pa F}{\pa g_{n+k}} .
\eeq
By introducing the resolvent 
$\hat{W}(z)=(1/N)\tr (z-\phi)^{-1}$,
eq.~(\ref{eqn:sd-eq}) is neatly summarized into a single
equality called the loop equation${}^1$
\begin{equation}
\la \hat{W}(z) \ra^2 -V'(z) \la \hat{W}(z)\ra +Q(z)=0,
\quad 
  Q(z)
\equiv 
  \sum_{k=1}^{m-1} \frac{V^{(k+1)}(z)}{k!}
 \left\langle \frac{1}{N} \tr (\phi-z)^{k-1} \right\rangle.
    \label{eqn:mat-q-function}
\end{equation}
Eqs.~(\ref{eqn:sd-eq})--(\ref{eqn:mat-q-function}) hold for a generic 
polynomial potential $V(\phi)=\sum_{k=1}^m \frac{g_k}{k} \phi^k$;
for our case $m=3$, $Q(z)$ is given by
$Q(z)=1+gz-g^2 +3g^3 \pa F/\pa g$.
Consequently we obtain a differential equation 
obeyed by the free energy $F(N,g)$
of the one-matrix model with the cubic coupling,
\bea
 &\!\! &\!\! \left[ N\frac{\partial}{\partial N} + 2 \right] F(N,g)=
 G\left(g, \frac{\partial F}{\partial g}\right)  + \OO\lt \oon\rt,
\quad 
\bar{\alpha}(g,a)\equiv\langle \alpha_{\rm s}\rangle
=-g+3g^2 a ,\nonumber \\  
  &\!\!  &\!\! G(g,a)=
   - \frac{g}{2}  a
   + \frac12 \bar{\alpha}^2
   + \frac g3 \bar{\alpha}^3
   + \log \lt 1 + g \bar{\alpha}\rt +
      \int^{-1/g - \bar{\alpha}}_{-\infty} dz
   \left(\la \hat{W}(z,g,a) \ra - \frac{1}{z}\right). 
\label{eqn:mat-nrge}  
\end{eqnarray}

To establish an algorithm for fixed points and scaling exponents, 
we first concentrate on the leading part $F^0(g)$ of the free
energy in the $1/N$-expansion.  
It is easy to see that $F^0$ satisfies
$  2 F^0(g) = G\left(g, \partial F^0(g)/\partial g \right)$.
We assume that $F^0$ consists of regular and singular parts 
around a fixed point $g_*$,
$  F^0(g) = \sum_{k=0}^\infty a_k(g-g_*)^k +
           \sum_{k=0}^\infty b_k(g-g_*)^{k+2-\gamma_0} \ \ \ 
(2-\gamma_0\not\in\N)$,
and that $G(g,a)$ is regular around $(g_*,a_1)$.
$\gamma_0$ is referred to as the susceptibility exponent.
By comparing the coefficients of
various powers of $g-g_*$ on both sides of the RG equation,
we have a set of equations to determine unknowns
$\gamma_0,\ g_*, \ a_k$ and $b_k$. 
We can also derive the RG equation for higher genus contributions 
and find that they exhibit the double scaling behavior
under a plausible assumption. 

We can as well use the eigenvalue representation
for the RG equation which is particularly useful for multi-coupling cases. 
We find four fixed points (Yang-Lee edge singularity, 
pure gravity and Gaussian) in the 
two-coupling ($g_3$ and $g_4$) case${}^6$;
there the spectrum of relevant operators of 
the exact solution 
($\gamma_0=-1/3,-3/2$ for Yang-Lee and $\gamma_0=-1/2$ for pure gravity)
are exactly reproduced.\\

\noindent
{\bf 3. The two-matrix model}\\
The partition function for the two-matrix model with 
$V(\phi,g)= \frac12 \phi^2 + \frac g3 \phi^3$ is defined as 
\begin{equation}
  Z_N(g_+,g_-,c) = 
\int d \phip d \phim
 \exp \left[-N  \tr ( V(\phip,g_+) + V(\phim,g_-) + c\phip \phim)\right].
\end{equation}
Similarly to the one-matrix case, we integrate the $N+1$-th row 
and column exactly and evaluate the singlet 
integral by the saddle point method. 
We consider the
case $g_+=g_-\equiv g$ which describes 
the Ising model coupled to 2D QG in
the absence of an external magnetic field.
We assume that 
$  \langle \tr \phip^j \rangle = \langle \tr \phim^j \rangle
  \label{eqn:symmetric-1pt-fn} $
holds for any $j$, and 
solve the reparametrization identities 
(a fourth-order algebraic equation for
the resolvent in need,
$\hat{W}_0(z)=
\oon {\rm Tr}
{\small 
 \left(
   \begin{array}{cc}
   z+\phip & c/g     \\
   c/g     & z+\phim
   \end{array}
 \right)^{-1} }$). 
Then we find 
a nonlinear RG equation for the two-matrix model
\begin{eqnarray}
\left[ N \frac{\partial}{\partial N} + 2 \right]F  
&\!\! =&\!\! G\left(g,c; 
     \frac{\partial F}{\partial g},
     \frac{\partial F}{\partial c}\right)
     + \OO\lt\oon\rt ,\nonumber \\
 G(g,c;a_g,a_c) 
&\!\! = \!\!  &
-  \frac{g}{2} a_g
+ ( 1 + c) \bar{\alpha}^2
         + \frac{2g}{3} \bar{\alpha}^3
+ 2 \log ( 1 + g \bar{\alpha}) \nonumber \\
& &   + \int_{- \infty}^{1/g+ \bar{\alpha}} dz \ 
     \left( \left\langle \hat{W}_0(z;g,c;a_g,a_c)\right\rangle
 - \frac 2z \right),
\nonumber \\
\bar{\alpha}(g,c;a_g,a_c)  &\!\! \equiv \!\!  &
\langle \alpha_{\pm, {\rm s}} \rangle=
\frac{g}{1+c} \lt -1+c \ a_c +\frac{3g}{2} a_g \rt 
\end{eqnarray}
We find three fixed points (critical Ising, 
pure gravity and Gaussian) 
in the physical region of the
coupling constant space${}^7$;
there the spectrum of relevant operators of 
the exact solution 
($\gamma_0=-1/3,-5$ for critical Ising and $\gamma_0=-1/2$ for pure gravity)
are again exactly reproduced.\\

\noindent
{\bf 4. Linearized RG flow}\\
We observe generically that the coefficients of the
nonlinear terms in the $G$-function, 
$\pa^n G/\pa a^n (g,0)$ for $n\geq 2$, are
suppressed by increasing powers of $g$.
Therefore we expect that within a region $g\ll 1$ where most
fixed points lie, the RG flow should be
well approximated by the linearized $\beta$-function,
$\beta_{\rm lin}(g)\equiv \pa G/\pa a (g,0)$.
Here we exhibit the RG flow in one- (Fig.1) and two-matrix models (Fig.2)
described by the real parts of $\beta_{\rm lin}$${}^7$.
We find that the critical lines (real lines) emanating from
the multicritical fixed points toward the pure gravity fixed point
are characterized as the RG trajectories corresponding to
the least relevant perturbations of the UV theories.
These flows are consistent with those for matter fields over
a fixed background, and suggest a gravitational analogue of 
the celebrated $c$-theorem${}^8$ stating the decrement of
the degrees of freedom of the matter sector along the RG flow.\\

\noindent
{\bf Acknowledgments}\\
N.S.\ thanks Institute for Theoretical Physics 
at Santa Barbara and the Aspen Center for Physics for hospitality 
where part of this article was written. 
This work is supported in part by 
Grant-in-Aid for Scientific Research (S.H.) and
(N.S., No.05640334), Grant-in-Aid for Scientific Research
for Priority Areas (N.S., No.05230019) from the Ministry of
Education, Science and Culture.\\

\noindent
{\bf References}\\
1. For reviews, see 
P.~Ginsparg and G.~Moore, 
{\em "Lectures on 2D gravity and 2D string theory"},\\
$\mbox{\ \ \ \ }$YCTP-P23-92, LA-UR-92-3479, hep-th/9304011
and references therein.\\
2. J.~Carlson,
{\em Nucl.~Phys.} {\bf B248} (1984) 536. \\
3. E.~{Br\'{e}zin} and J.~{Zinn-Justin},
{\em Phys.~Lett.} {\bf B288} (1992) 54.\\
4. V.~Periwal,
{\em Phys.~Lett.} {\bf B294} (1992) 49;
J.~Alfaro and P.~Damgaard,
{\em ibid.} {\bf B289} (1992) 342;\\
$\mbox{\ \ \ \ }$C.~Ayala,
{\em ibid.} {\bf B311} (1993) 55;
Y.~Itoh, {\em Mod.~Phys.~Lett.} {\bf A8} (1993) 3273.\\
5. S.~Higuchi, C.~Itoi and N.~Sakai,
{\em Phys.~Lett.} {\bf B312} (1993) 88;\\
$\mbox{\ \ \ \ }$S.~Higuchi, C.~Itoi and N.~Sakai,
{\em Prog.~Theor.~Phys.~Suppl.} {\bf 114} (1993) 53.\\
6. S.~Higuchi, C.~Itoi, S.~Nishigaki and N.~Sakai,
{\em Phys.~Lett.} {\bf B318} (1993) 63.\\
7. S.~Higuchi, C.~Itoi, S.~Nishigaki and N.~Sakai,
{\em "Renormalization group flow in 
one- and \\
$\mbox{\ \ \ \ }$two-matrix models"},
TIT/HEP-261, 
NUP-A-94-16 (Sept. 1994), hep-th/9409009.\\
8. A.~B.~Zamolodchikov,
{\em JETP Lett.} {\bf 43} (1986) 730.\\

\ \\

\noindent
\begin{minipage}{\textwidth}
\begin{center}
    \leavevmode
    \figinclude{237}{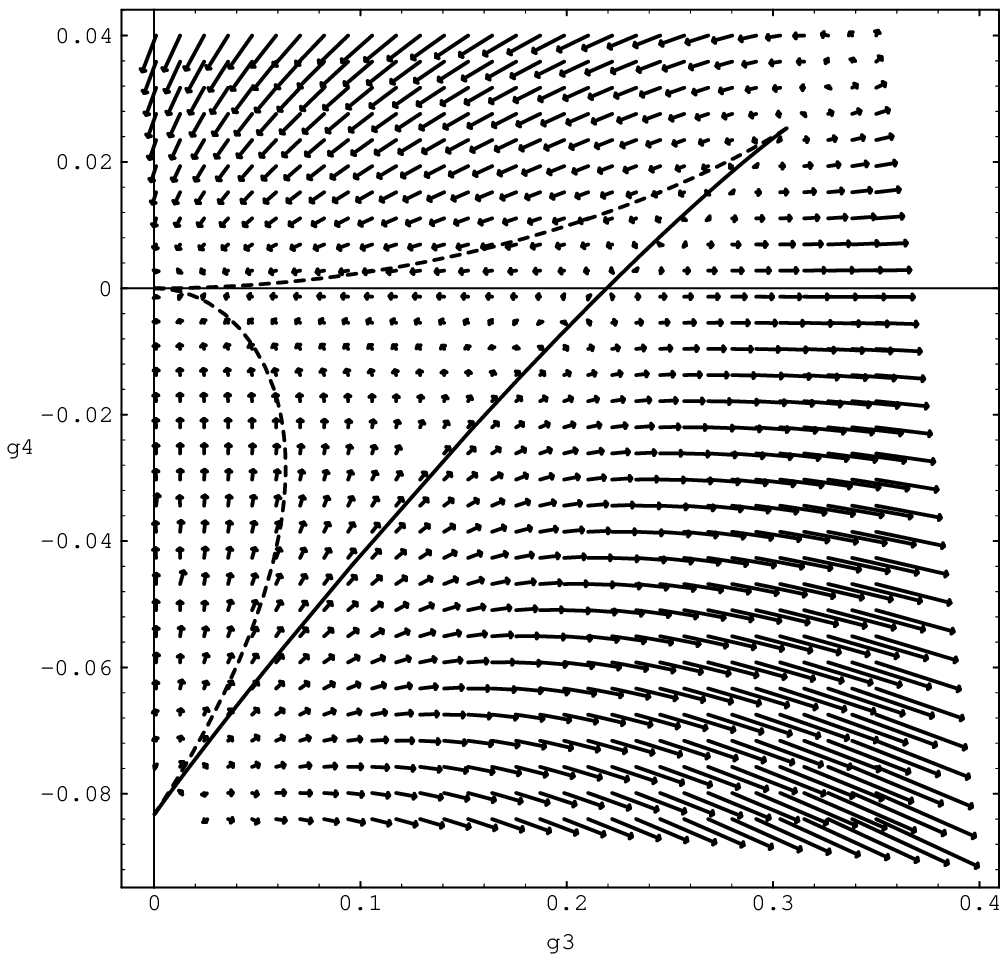} \ \ \ \ 
 \figinclude{237}{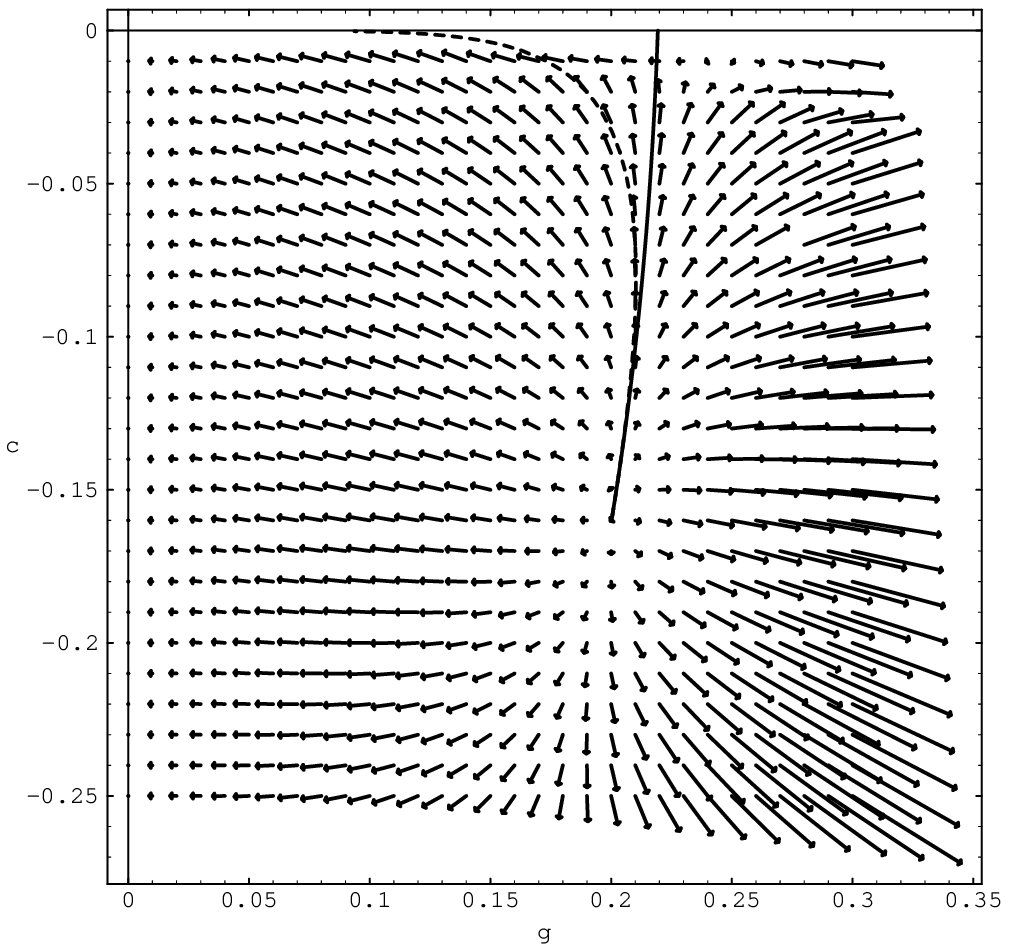}\\

\medskip
\noindent 
{\small 
\ \ \ \ \ Figure 1: RG flow for the one-matrix model.
\ \ \ \ \ \ \ \ \ \ 
Figure 2: RG flow for the two-matrix model.\\ 
The arrows represent a flow from the UV $(N=\infty)$ to the IR $(N=0)$.}
\end{center}
\end{minipage}

\end{document}